\documentclass[12pt,semidraft,epsfig]{article}
\usepackage{graphicx}

\def\s#1{\setbox0=\hbox{$#1$}%
\rlap{\ifdim\wd0>.7em\kern.22\wd0\else\kern.1\wd0\fi /}#1}

\newcommand{\beq}{\begin{equation}}
\newcommand{\eeq}{\end{equation}}
\newcommand{\bea}{\begin{eqnarray}}
\newcommand{\eea}{\end{eqnarray}}

\newcommand{\tto}{\!\to\!}

\newcommand{\gsim}{\lower.7ex\hbox{$
\;\stackrel{\textstyle>}{\sim}\;$}}
\newcommand{\lsim}{\lower.7ex\hbox{$
\;\stackrel{\textstyle<}{\sim}\;$}}

\newcommand{\bibit}[1]{\bibitem{#1}}

\newcommand{\La}{\overline{\Lambda}}

\newcommand{\lab}{\Lambda_b}

\newcommand{\GeV}{\,\mbox{GeV}}
\newcommand{\MeV}{\,\mbox{MeV}}
\newcommand{\matel}[3]{\langle #1|#2|#3\rangle}

\newcommand{\eod}{\end{document}}

\newcommand{\msp}[1]{\mbox{\hspace*{#1mm}~}}

\begin{document}
\thispagestyle{empty}
\vspace*{-40pt}

\begin{flushright}
UND-HEP-11-BIG\hspace*{.08em}02\\

\end{flushright}
\vspace*{35pt}

\begin{center}
\scalebox{1.1}{\Large {\bf \hspace*{-1pt}Semileptonic width ratios among 
beauty hadrons}}\vspace*{50pt}

{\large \bf I.I.~Bigi$,^a\;\;$ Th.~Mannel$,^b\;\;$  N.~Uraltsev$\,^{a,b,c}$}\\
\vspace{20pt}\small 
$^a$ {\sl Department of Physics, University of Notre Dame du Lac, 
Notre Dame, IN 46556, USA}\\[3pt]
{\sl $^b$ Theoretische Physik 1, Fachbereich Physik,
Universit\"at Siegen\\ D-57068 Siegen, Germany}\\[2pt] 
{\sl  $^c$ St.\,Petersburg Nuclear Physics Institute, Gatchina,
St.\,Petersburg 188300, Russia}\\

\normalsize

\vspace*{60pt}

{\bf Abstract}\vspace{-1.5pt}\\
\end{center}

\noindent
We present predictions 
based on the heavy quark expansion in QCD. We find $SU(3)$ breaking in
$B$ mesons suppressed in the framework of the HQE. $B_s$ is expected to have
the semileptonic width about $1\%$ lower and  $\Lambda_b$ about $3\%$ higher when 
compared to $\Gamma_{\rm sl}(B_d)$.  
The largest partial-rate preasymptotic effect is Pauli interference in the
$b\tto u\, \ell\nu$ channel in $\Lambda_b$, about $+10\%$.
We point out that the $\Omega_b$ semileptonic
width is expected not to exceed that of $B_d$ and may turn out to be the
smallest among stable $b$ hadrons despite the large mass. The
underlying differences with phase-space models are briefly addressed
through the heavy mass expansion.

\newpage

\tableofcontents
\vspace{15pt}

\section{Introduction}
\label{intro}

At the LHC a new dedicated flavor experiment has come into operation, namely
the LHCb. Big data sets of $B_u$ and $B_d$ transitions have been investigated
at the $B$ factories by Belle and BaBar and at FNAL by CDF and D0; smaller
data sets for $B_s$ and $\Lambda_b$ decays have been studied by CDF and D0
(and also by Belle for $B_s$). LHCb will generate even much larger sets of
weakly decaying beauty mesons and baryons, including $\Xi_b$ and $\Omega_b$.
The lifetimes of $\Lambda_b$, $\Xi_b$ and $\Omega_b$ will be well measured as
will be the $B_s$ total width and $\Delta \Gamma_{\!B_s}$. LHCb will also
analyze semileptonic channels. The future Super-Flavor Factories will
measure inclusive semileptonic rates for these $b$ hadrons with good
accuracy. The anticipated experimental precisions should be matched with
reliable theoretical predictions incorporating nonperturbative effects.

In a recent paper \cite{gr} a question was raised about the difference between
the inclusive semileptonic decay rates of different heavy flavor hadrons;
beauty particles represent the most interesting case in this respect. It is
appropriate to summarize the up-to-date predictions of the existing
QCD-based theory. In retrospect, the numerical aspects of the
predictions for the differences in the semileptonic widths derived from the
Heavy Quark Expansion (HQE) have been addressed so far occasionally
\cite{WADs,vub}, with more emphasis on the related theoretical aspects or as a
supplementary tool to other studies.  The main attention has been
dedicated to the lifetimes of beauty particles, even though it had been
appreciated from the earliest days of the HQE that the semileptonic decays
generally allow more precise predictions.

We find the HQE predictions for the inclusive beauty semileptonic decay rates differ
significantly from those in Ref.~\cite{gr} which are mostly based on a
simple phase-space model. An experimental verification would serve as yet
another instructive example of the role of the consistent treatment of
the nonperturbative strong dynamics. We also briefly consider
the major differences with the naive models in the context of the
heavy mass expansion.

\section{Heavy quark expansion with $SU(3)$ breaking}

The treatment of the nonperturbative QCD effects on the fully inclusive decay rates
of the heavy flavor hadrons is provided by the OPE-based HQE
\cite{versus,bs}.\footnote{The idea that the total weak decay rates 
of heavy flavor hadrons asymptotically approach  
the free-quark rate goes back to the paper \cite{nnnikol}.}
Its two principal points in the present context are:  
\begin{itemize}
\item
The overall decay probability does not have corrections linear in $1/m_Q$. 
In other words, the mass itself of the decaying hadron
does not affect the decay rate; the rate is rather determined by the (short-distance)
masses of the quarks appearing in the weak decay Lagrangian. Hadron masses in the
final state are only indirectly related to the width through various sum rules
constraining the relevant nonperturbative QCD expectation values in the
decaying heavy flavor hadron.
\item 
The tower of the local heavy quark operators and their coefficients for a
given quark-level channel appearing in the $1/m_Q$-expansion of the widths
are universal.  This means in turn, that all the dependence on a decaying hadron
lies in the expectation values of these universal series of operators. This
feature was not self-manifest for the historically first considered 
nonperturbative effects, namely weak annihilation (WA) and Pauli interference
(PI) \cite{wapi},
in particular when introduced through simple quark diagrams. 
It was treated properly later \cite{mirage,WADs}.
\end{itemize}

\noindent
As a consequence, the analysis of the $SU(3)$-breaking pattern in
$\Gamma_{\rm sl}(B_s)$ vs.\ $\Gamma_{\rm sl}(B_d)$ requires the estimates of
the difference in the expectation values of the leading heavy quark 
operators between $B_s$ and $B$ in the series \cite{imprec}  
\bea
\nonumber
\Gamma_{\rm sl}(B)\!=\! \frac{G_F^2}{192\pi^3}|V_{cb}|^2 m_b^5 
z_0(\mbox{$\frac{m_c^2}{m_b^2}$}) \msp{-5}& &\msp{-5}
\left[
1+ c_\pi \frac{1}{2m_b^2}\frac{\matel{B}{\bar{b}(i\vec D)^2b}{B}}{2M_B} 
+ c_G
\frac{1}{2m_b^2}
\frac{\matel{B}{\bar{b}(\frac{i}{2}\sigma_{\mu\nu}G^{\mu\nu}b}{B}}{2M_B}
+ \right. \\
& &\msp{-40} \left. c_D \frac{1}{m_b^3}
\frac{\matel{B}{\bar{b}(\mbox{$-\frac{1}{2}$}\vec D \vec E)
    b}{B}}{2M_B} + 
\frac{32\pi^2}{z_0 m_b^3}\frac{\matel{B}{\bar{b}\vec \gamma(1\!-\!\gamma_5)c\,
\bar c\vec \gamma(1\!-\!\gamma_5) b}{B}_{\rm IC}}{2M_B}+ ...
\right]\!,
\label{30}
\eea
where the ellipses stand for the
higher orders in $1/m_b$. 

The last term explicitly given in Eq.~(\ref{30}) describes the subset of the
higher-order nonperturbative corrections generically referred to as `Intrinsic
charm' (IC) -- a term not fully adequate here from a theoretical perspective.
Various corrections to the Wilson coefficients -- including the often
technically challenging perturbative renormalization important for precision
evaluation of the inclusive widths -- are not mandatory here.  Nor is
generally a scrupulous treatment of the renormalization point in the heavy
quark operators, as long as it is of a reasonable scale.  Likewise, accounting
for the significant $\tau$-lepton mass is not critical unless a precise
prediction specifically for the channel $b\tto c(u)\,\tau\nu_\tau$
\cite{koytau} is aimed at.  For the decay width mediated by the $b\tto
u\,\ell\nu$ transitions the four-fermion `WA' operator $\bar b u\, \bar u b$
is required replacing the `IC' expectation value, which is not a $SU(3)$ or
$V$-spin singlet.

A summary of the breakdown of the  estimated power corrections in the
semileptonic $B$ decays can give a starting idea about the expected
effects \cite{hiord}: 
\bea
\nonumber \frac{1}{m_b^2}\,: \msp{-4} & & \msp{-1} 
\frac{\delta_{\mu_\pi^2}\Gamma_{\rm sl}}{\Gamma_{\rm sl}} \approx -1\%, \qquad
\frac{\delta_{\mu_G^2}\Gamma_{\rm sl}}{\Gamma_{\rm sl}} \approx -3.5\% \\
\nonumber \frac{1}{m_b^3}\,: \msp{-4} & & \msp{-1} 
\frac{\delta_{\rho_D^3}\Gamma_{\rm sl}}{\Gamma_{\rm sl}} \approx -3\% \\
\nonumber \frac{1}{m_b^4}\,: \msp{-4} & & \msp{-1} 
\frac{\delta \Gamma_{\rm sl}}{\Gamma_{\rm sl}} \approx 0.5\% \\
\frac{1}{m_Q^5}\,: \msp{-4} & & \msp{-1} 
\frac{\delta_{\rm IC} \Gamma_{\rm sl}}{\Gamma_{\rm sl}} \approx 0.7\%, \qquad \;
\frac{\delta_{\rm tot} \Gamma_{\rm sl}}{\Gamma_{\rm sl}} \approx 0.5\%\,;
\label{36}
\eea 
these estimates were derived for non-strange $B$ mesons. 
The spin-orbit expectation
value $\rho_{LS}^3$ emerging at the $1/m_b^3$ level enters the width only as a
part of the full Lorentz-invariant quantity $\mu_G^2$; its effect in 
the differential distributions is typically insignificant. 

An important feature is the smallish impact of the kinetic operator. Its
coefficient $c_\pi\!=\!-1$ is universally suppressed in the integrated 
rates \cite{spectra}; in a certain sense it can even be regarded as vanishing. 
Therefore a realistic variation of $\mu_\pi^2$ will not lead to a 
relevant direct effect on
$\Gamma_{\rm sl}(B_s)/\Gamma_{\rm sl}(B_d)$. We thus see that the main effect
potentially comes from the chromomagnetic interaction or from the Darwin
term. Assuming the typical scale of about $30\%$ for the $SU(3)$ violation in the 
expectation values we may
a priori expect a difference of around two percent in $\Gamma_{\rm sl}(B_s)$. 
Our actual prediction turns out lower and definitely favors the
suppression of the $B_s$ width.

We present a more detailed discussion of the $SU(3)$ breaking effects in the
following. The primary inputs on the experimental side are the masses of $B$,
$B^*$ and $D$, $D^*$, both strange and non-strange. 
To build a consistent physical picture of the effects we often
confront our expectations qualitatively to the explicit pattern of the heavy
quark expansion numerically studied in Ref.~\cite{lebur} in the context of the
exactly solvable 't~Hooft model \cite{thooft}, a large-$N_c$ limit of QCD in
two dimensions.

\subsection{$\mu_G^2$}

The chromomagnetic expectation value $\mu_G^2$ is most directly extracted from
the hyperfine splitting $M_{B^*}-M_B$:
\beq
M_{B^*}\!-\! M_B \simeq \frac{4}{3} \,\tilde c_G \,\frac{\mu_G^2}{2m_b}, 
\qquad \tilde c_G \approx 1.
\label{44}
\eeq
Using  $M_{B^*}\!-\!M_B\!=\!45.78 \pm 0.35\MeV$,  
$M_{B_s^*}\!-\!M_{B_s}\!=\!49.0 \pm 1.5\MeV$ we arrive at a tiny $SU(3)$
violation in the hyperfine splitting:
\beq
\frac{\mu_G^2(B_s)}{\mu_G^2(B_d)} \simeq 1.07\pm 0.03 \; ;
\label{46}
\eeq
qualitatively it can be regarded as nearly absent. In principle, the
$1/m_b$ corrections to the relation (\ref{44}) are about $15\%$ \cite{chrom}.
However, they are smaller in the ratio; this is supported by the similar
equality in the $D$ system:
\beq
M_{D_s^*}\!-\!M_{D_s}=  143.8 \pm 0.4\MeV \mbox{~~vs.\ ~} M_{D^{+*}}\!-\!M_{D^+}= 
140.65 \pm 0.1\MeV .
\label{48}
\eeq

Taken at face value the $SU(3)$ breaking (\ref{46}) would lead to a small
shift
\beq
\delta_{\mu_G^2}\,\frac{\Gamma_{\rm sl}(B_s)}{\Gamma_{\rm sl}(B_d)} 
\simeq -0.25\%.
\label{50}
\eeq
The actual uncertainty is probably at least two thirds of the value itself. 

\subsection{$\mu_\pi^2$}

The overall impact of the kinetic operator is a few times smaller than the
chromomagnetic effect; therefore no high precision in it is required. Its
$SU(3)$-nonsinglet component can be estimated comparing the observed mass
shifts in $B$ and $D$ systems \cite{versus}; one can either use the 
spin-averaged combinations or just the pseudoscalar masses proper, since the
hyperfine splitting turns out nearly $SU(3)$-blind. 
We have 
\bea
\mbox{$\frac{3M_{B_s^*}\!+\!M_{B_s}}{4}\!-\!\frac{3M_{B^*}\!+\!M_{B}}{4}$}
\!=\!(86.8\!+\!2.4)\MeV 
\msp{-5}& \mbox{ vs.\ }& \msp{-5} 
\mbox{$\frac{3M_{D_s^*}\!+\!M_{D_s}}{4}\!-\!\frac{3M_{D^*}\!+\!M_{D}}{4}$}\!=\!
(98.9\!+\!2.6)\MeV; \rule[-10pt]{0pt}{15pt} \qquad\label{52a} \\
M_{B_s}-M_{B}= 86.8 \MeV 
& \mbox{~ vs.\ ~}&  M_{D_s}-M_{D}=
(98.9 \pm 0.3)\MeV \; .
\label{52b}
\eea
The differences amount to $\La_s-\La$, which is therefore about $85\MeV$. The
variation in the estimate between beauty and charm is an effect of $1/m_Q$
(and higher) terms. Neglecting the higher-order corrections and using
$m_b=4.6\GeV$, $m_c=1.25\GeV$  we get
\beq
\mu_\pi^2(B_s)-\mu_\pi^2(B_d)\simeq 0.041\GeV^2
\label{54}
\eeq
in either way. (Certain allowance should be left for the electromagnetic
effects.) This means a $10\%$ $SU(3)$ breaking in
$\mu_\pi^2$ and only a per mil stronger suppression of 
$\Gamma_{\rm sl}(B_s)$ vs. $\Gamma_{\rm sl}(B_d)$.

One should, however, consider that the $1/m_Q^2$ corrections in the meson
masses are as large as $30$ to $50\MeV$ in charm \cite{f0short}. 
Even though their $SU(3)$
symmetry softens the impact on the splitting in question, the quality of the
$SU(3)$ relations generally deteriorates for higher-order expectation
values. To account for this we shall adopt an interval of larger $SU(3)$
breaking; our expectations actually center about
the twice larger effect than in Eq.~(\ref{54}):  
\beq
\mu_\pi^2(B_s)-\mu_\pi^2(B_d)\simeq 0.08\; \mbox{to} \: 0.1 \GeV^2 \; ;
\label{56}
\eeq
i.e.\ up to $25\%$. This does not really affect directly 
$\Gamma_{\rm sl}(B_s)$ at an appreciable level; 
it refers more to gaining a self-consistent dynamic picture of the heavy mesons. 

\subsection{$\rho_D^3$}

The Darwin expectation value $\rho_D^3$ from dimension-six operators for $B$
mesons is a priori less certain -- as expected when the dimension of the
operators is increased. A reasonable estimate for it is provided by the exact
sum rules applicable in the heavy quark limit. Assuming their dominance by the
lowest $P$-wave states with $\bar\epsilon_P\!\simeq\! 400\,\mbox{to}\,450\MeV$
we obtain
\beq
\rho_D^3 \simeq \bar\epsilon_P \mu_\pi^2 \simeq 0.18\GeV^3, \qquad
\rho_D^3 \gsim \frac{2}{3} \frac{(\mu_\pi^2)^2}{\La} \simeq 0.18\GeV^3 
\label{58}
\eeq
apparently well fitting the data. Blindly using the same relation in $B_d$
and in $B_s$ for comparing $\rho_D^3$ and $\rho_D^3(B_s)$ we would have gotten
\beq
\frac{\rho_D^3(B_s)}{\rho_D^3} \simeq 
\left(\frac{\mu_\pi^2(B_s)}{\mu_\pi^2}\right)^2 \,\frac{\La\,}{\La_s\!} \quad 
\mbox{with ~} \La_s\!-\!\La \simeq 82\MeV ,
\label{60}
\eeq
and thus a ratio about $1.27$  assuming a $20\%$ increase in
$\mu_\pi^2(B_s)$ compared to $\mu_\pi^2(B_d)$, see Eq.~(\ref{56}). (This ratio
becomes $1.07$ if using the number from  Eq.~(\ref{54}).) 
However, the validity of the estimates like Eq.~(\ref{58}) may not necessarily be
equal for different spectator mass in $B$ meson; in other words, even a fair
overall accuracy of such relations does not automatically apply 
to their derivative in respect to spectator mass $m_s$ as would be 
required to justify Eq.~(\ref{60}).\footnote{We did observe this to hold in 
the 't~Hooft model when studied the effects of duality violation in its setting.}

An alternative evaluation of $\rho_D^3$ dating back to 1993 \cite{motion}
relies on the vacuum factorization estimate for the four-fermion heavy quark
operator the Darwin operator reduces to by the gauge field equations of
motion:
\beq
\rho_D^3 \approx \frac{g_s^2}{18} f_B^2 M_B  .
\label{62}
\eeq
Since the factorized contribution is leading in $N_c$ for the Darwin operator,
it is a reasonable 
guess.\footnote{The situation is different in this respect for
the typical four-quark operators encountered in the analysis of the
flavor-dependent corrections in the lifetimes.} For the ratio the precise
scale entering the strong coupling does not matter, and we get
\beq
\frac{\rho_D^3(B_s)}{\rho_D^3} \approx \frac{f_{B_s}^2}{f_B^2}\,.
\label{64}
\eeq
Such a relation is exact in the heavy quark limit in the 't~Hooft model
\cite{d2,burkur}. 

The actual values of the axial decay constants in $B$ mesons are not yet well
known, although the question has gotten much attention in the past years. It
was suggested in the 1980s \cite{blinov} that
\beq
\frac{f_{B_s}}{f_B} \gsim  \frac{f_K}{f_\pi} \, ; 
\label{66}
\eeq
this leads to $f_{B_s}^2/f_B^2 \approx 1.4 \div 1.7$, a ratio
preferred nowadays. 
Regardless of a concrete reasoning, it is more than
just plausible that $f_B$ -- a wavefunction density at origin -- increases with
the mass of the light quark in the meson for both heavy-light and light-light
bound states. 

Various arguments lead us to suggest that $\rho_D^3(B_s)/\rho_D^3$ is larger
than unity; the question is rather by how much $SU(3)$ is violated in this 
parameter. We think that a priori the natural scale is $40$ to $50\%$. On the
other hand, the heavy quark relations seem to favor somewhat softer $SU(3)$
breaking effects around $20\%$. Since both lines of reasoning are quite
general although more qualitative,  the most natural scenario seems to lie in
between:
\beq
\frac{\rho_D^3(B_s)}{\rho_D^3} \approx 1.25  .
\label{70}
\eeq
With the significant negative coefficient $c_D\!\approx\!-16$ \cite{grekap} this  
yields an additional suppression of $\Gamma_{\rm sl}(B_s)$ by about
$0.8\%$.

\subsection{Higher orders in $1/m_b$}

The number of operators in the series for $\Gamma_{\rm sl}(B)$ 
proliferates fast in higher orders. Moreover, the $SU(3)$ symmetry in their
individual expectation values can be strongly violated being restored to
the `normal' level only in the aggregate effect on the observables. Therefore,
a detailed analysis paralleling the one given above would become less and less  
meaningful. A more reasonable perspective is to calibrate the
overall effect and to assume an up to $50\%$ breaking of $SU(3)$ in higher orders. 
Furthermore, we would generally expect their enhancement in $B_s$ vs.\
$B_d$; this is justified provided the net effect does not come as a result of
significant cancellations.

The recent analyses estimated the effects of higher-order OPE terms at the one 
percent level being dominated by the so called IC corrections
\cite{ic,icsieg,hiord}:  
\beq
\frac{\delta_{1/m_b^4}\Gamma_{\rm sl}(B)}{\Gamma_{\rm sl}(B)} \approx 0.5\%,
\qquad \frac{\delta_{\rm IC}\Gamma_{\rm sl}(B)}{\Gamma_{\rm sl}(B)} \approx
0.7\%, \qquad \frac{\delta^{\rm tot}_{1/m_b^5}
\Gamma_{\rm sl}(B)}{\Gamma_{\rm sl}(B)} \approx 0.5\%.
\label{72}
\eeq
Based on the above prescription we expect
\beq
\frac{\delta^{\rm hi\,ord}\Gamma_{\rm sl}(B_s)\!-\!
\delta^{\rm hi\,ord}\Gamma_{\rm sl}(B)}{\Gamma_{\rm sl}(B)} \lsim 0.5\% \; .
\label{74}
\eeq

{\sl To summarize:} We expect the dominant $SU(3)$-breaking
correction to 
$\Gamma_{\rm sl}(B_s)/\Gamma_{\rm sl}(B_d)$ to come from the Darwin operator
constituting an impact of about $-0.8\%$, followed by the smaller negative
effects from the chromomagnetic and kinetic operators with $-0.25\%$ and
$-0.15\%$, respectively; the latter two 
could be offset by a small
relative enhancement of the $B_s$ width by about $0.5\%$ from higher-order
power corrections.

\subsection{$SU(3)$ breaking and heavy quark sum rules}
\label{sumrules}

It is instructive to recall that the leading heavy quark nonperturbative
parameters are given by the moments of the positive SV structure functions of
the respective $B$ mesons:
\beq
\varrho^2\!-\mbox{$\frac{1}{4}$} \!=\!\! \int \! W_+(\varepsilon) \, 
{\rm d}\varepsilon, \;\;\;
\La\!=\! 2\! \!\int \! W_+(\varepsilon) \, 
\varepsilon {\rm d}\varepsilon, \;\;\;
\mu_\pi^2\!=\! 3\!\!\int \! W_+(\varepsilon) \, 
\varepsilon^2 {\rm d}\varepsilon, \;\;\;
\rho_D^3\!=\! 3\!\!\int \! W_+(\varepsilon) \, 
\varepsilon^3 {\rm d}\varepsilon.
\label{80}
\eeq
For the $SU(3)$ breaking the difference $W_+^{(B_s)}\!-\!W_+^{(B_d)}$ enters.
Of course, the moment relations are less constraining when the function is no
longer positive. Nevertheless certain qualitative conclusions can be drawn.

We have observed some increase in the relative $SU(3)$ breaking for
higher moments of $W_+(\varepsilon)$ -- from the first ($\La$) to the third 
($\rho_D^3$), which looks natural. Extrapolating this to the zeroth moment we
would arrive to expect a small increase in the IW slope in $B_s$ compared to
$B_d$. 

Let us note that the particular $SU(3)$-breaking pattern
\beq
\La_s\!-\!\La\!=\!82\MeV, \quad
\mu_\pi^2(B_s)\!-\!\mu_\pi^2(B_d)\!=\!0.06\GeV^2, \quad 
\rho_D^3(B_s)\!-\!\rho_D^3(B_d)\!=\!0.04\GeV^3 
\label{82}
\eeq
would be well described, for $\La$ and $\mu_\pi^2$, by a small increase in
$2|\tau_{3/2}|^2+|\tau_{1/2}|^2$ by $0.1$ at $\varepsilon\!\simeq\!0.45\GeV$; 
yet it would be somewhat worse for $\rho_D^3$ where additional
contributions from higher strange $P$-waves are then preferred. This
qualitative solution (not unique, of course) would imply the increase in the
strange IW slope by about $0.1$:
\beq
\varrho^2_{B_s}-\varrho^2 \approx 0.1\,. 
\label{84}
\eeq
In the 't~Hooft model the overall $m_s$-dependence has been observed to
be weaker both in the leading heavy quark parameters and in the meson axial
constant; the dominance of the first $P$-wave excitation in the sum rules 
was excellent for either spectator mass. If we accept the same to be the 
case in actual QCD we would end up with a lower $SU(3)$ breaking in the Darwin value
and in the $f_{B_s}/f_B$ ratio than it has been estimated in
the previous sections.

We have found a further reduced $SU(3)$ breaking in $\mu_G^2$, which is smaller
than can be
expected from any sort of nonrelativistic quark model. This, however is more
natural for a highly relativistic bound state by virtue of the sum rules: 
the exact relation equating the zeroth moment of the $W_-(\varepsilon)$ SV
structure function to the spin of the light cloud 
\cite{newsr} mandates complete absence of
the $m_s$ effects in this moment, a property apparently reflected, in a weaker
form, in the second moment yielding $\mu_G^2$. 

Let us note in passing that in the BPS limit \cite{BPS} 
$\mu_\pi^2\!=\!\mu_G^2$ for
non-strange $B$ mesons the first-order  $SU(3)$-breaking corrections in the
difference $\mu_\pi^2\!-\!\mu_G^2$ (as well as in $\varrho^2$) must vanish 
starting with terms $\propto \! m_s^2$. The proximity to the BPS
regime may therefore explain the apparent smallness of the relative 
$SU(3)$-breaking in the lowest-dimension nonperturbative heavy quark
parameters.

\section{The heavy quark expansion for $\Gamma_{\rm sl}(\Lambda_b)$}

No unitary symmetry per se relates the decay rates of beauty baryons and
mesons; the corresponding expectation values must be analyzed separately. 

Since the light degrees of freedom in $\Lambda_b$ carry no spin in the heavy quark
limit, the structure of the $1/m_Q$ series simplify compared to $B$ mesons:
only spin-singlet expectation values survive. In particular, the contribution
from the chromomagnetic operator vanishes to the leading order in $1/m_b$.

\subsection{$\mu_\pi^2$}

As suggested in Ref.~\cite{versus}, we have estimated 
$\mu_\pi^2(\lab)\!-\!\mu_\pi^2$ following the analogue of Eq.~(\ref{52b}) via
\bea
\nonumber
\La_\Lambda-\La \msp{-2} & \simeq & \msp{-2} M_{\lab}-\bar M_B\approx 305\MeV\\
\left( \!\frac{1}{2m_c}\!-\!\frac{1}{2m_c}\!\right) 
[\mu_\pi^2(\lab)\!-\!\mu_\pi^2]+ {\cal O}\!
\left(\!\mbox{$\frac{1}{m_Q^2}$}\!\right)
\msp{-4} &
\simeq & \msp{-4} \left(M_{\Lambda_c}\!-\!\bar M_D\right) -
\left(M_{\lab}\!-\!\bar M_B\right) \simeq 5.4\MeV \qquad
\label{90}
\eea
which literally would yield $\mu_\pi^2(\lab)\!-\!\mu_\pi^2 \!\simeq\!
0.02\GeV^2$. However, higher-order corrections in $1/m_c$ are more significant. 
In particular, the $1/m_c^2$ terms were estimated
\cite{f0short} to be about $-45\MeV$ in $\bar M_D$, which by 
itself would
raise the $\mu_\pi^2$ difference by $0.15\GeV^2$. Little is known 
at the moment about them
in $\Lambda_c$; they can generally be expected similar in size. We thus estimate
\beq
\mu_\pi^2(\lab)\!-\!\mu_\pi^2\approx 0.1\pm 0.1 \GeV^2;
\label{92}
\eeq
the uncertainty figure here is not iron-clad, yet the precise value is not
of a direct importance for the width difference. Therefore we conclude that
\beq
\delta_{\mu_\pi^2} \,\frac{\Gamma_{\rm sl} (\lab)}{\Gamma_{\rm sl}(B_d)} 
\approx -0.25\%\,.
\label{94}
\eeq

\subsection{$\rho_D^3$ and higher orders}

Little as well is known directly about the Darwin expectation value in
$\lab$. Its value was estimated in Ref.~\cite{four} to be about $0.15\GeV^3$ -- 
somewhat smaller than $\rho_D^3$ in $B$ mesons. The related estimate of the
kinetic expectation value in $\lab$ yielded the result quite close to
Eq.~(\ref{92}). Our expectation then is that 
\beq
\delta_{\rho_D^3} \,\frac{\Gamma_{\rm sl} (\lab)}{\Gamma_{\rm sl}(B_d)} 
\simeq (1\pm 1.5)\%\, ,
\label{96}
\eeq
tentatively the next-to-largest effect after the vanishing of the
chromomagnetic term. 

The question of the higher-order corrections is somewhat subtle. The analysis
of the leading IC-related contributions in $B$ mesons indicated \cite{ic} 
that the
nonperturbative corrections to the vector $\bar b b \, \bar c \gamma_0 c$ 
expectation value dominate, but are partially moderated by those in 
the axial piece given by $\bar b \vec \sigma b \, \bar c \vec \gamma c$. 
In $\lab$ the axial piece vanishes to the leading order since the light degrees
of freedom form a spinless state. The overall IC estimate \cite{ic,icsieg} can
readily be adapted to $\lab$ by setting $\mu_G^2=\rho_{LS}^3=0$ and using the
appropriate values for $\mu_\pi^2$, $\rho_D^3$ and $\bar\epsilon_P$. The
estimate of the two-loop IC effects $\propto \! \alpha_s/m_cm_b^3$ can
likewise be obtained with the suitable adaptation of the analysis in
Ref.~\cite{ic}. This results in a tentative estimate of a similar effect of
the same sign, yet somewhat suppressed by a factor of $0.5$ to $0.7$. These
considerations suggest that
\beq
\delta_{\rm IC}\, \frac{\Gamma_{\rm sl}(\lab)}{\Gamma_{\rm sl}(B)} \gsim -0.3\%.
\label{98}
\eeq

A similar analysis can be applied to the regular $1/m_b^4$ and $1/m_b^5$
corrections. We obtain a mild overall suppression of the
combined higher-order effect in $\lab$, by about $30\%$ of that in $B$ 
mesons, thus
mirroring the numerics in Eq.~(\ref{98}). This estimate in only tentative,
though. 

It is interesting to note that the heavy quark expectation
values in $\lab$ do not appear to be remarkably larger than their $B$
counterparts: the kinetic value emerges close, and the Darwin value is
possibly even smaller than in $B$. This comes in contrast with $\La$ which is
significantly larger; it evidently scales like $N_c$ in heavy baryons. Through
the SV sum rules we conclude that the IW slope must be significantly larger in
$\lab$ (in particular, once the dynamically-generated slopes 
are counted on the same footing), the
fact almost inevitable in any model. The similarity of the dynamic properties
of $\lab$ and $B$ in the large-$N_c$ expansion has been emphasized by
M.~Shifman based on the orientifold approach to the large-$N_c$
QCD \cite{orient}.

{\sl To summarize:} Our expectations for $\Gamma_{\rm sl}(\lab)/\Gamma_{\rm sl}(B)
\!-\!1$ center around $+3\%$ being dominated by the absence of the
chromomagnetic suppression in $\lab$ and otherwise only slightly affected by
the corrections from the Darwin and kinetic operators and from higher-order
effects. 

\subsection{A note on $\Gamma_{\rm sl}(\Omega_b)$}

In a distant future the double-strange $\Omega_b$-baryon may represent an 
intriguing case for a precision test of the HQE predictions for the decay
rates. It is the only stable beauty baryon with a spin-$1$ light cloud
(its hyperfine twin $\Omega_b'$ decays electromagnetically), and
it is significantly heavier than either $B$ mesons or $\Lambda_b$, 
$M_{\Omega_b}\!\approx\!6.05\GeV$, being a separate light-flavor 
symmetry state. In spite of the large mass
its semileptonic width may not be any noticeably larger than for
$\Lambda_b$ or $B$ -- probably, it lies between the two. Based on the
measured hyperfine splitting in $\Omega_c$, the width is expected to
be suppressed by the chromomagnetic interaction at the level of $1.5\%$, 
\beq
 \frac{\delta_{\mu_G^2}\Gamma_{\rm sl} (\Omega_b)}{\Gamma_{\rm sl}(B_d)} 
\simeq \frac{8}{9}\frac{M_{\Omega_c^*}\!-\!M_{\Omega_c}}{M_{D^*}\!-\!M_D} 
\cdot \frac{\delta_{\mu_G^2}\Gamma_{\rm sl} (B_d)}{\Gamma_{\rm sl}(B_d)}
\approx -1.5\%\,,
\label{194}
\eeq
i.e.\ only about $45\%$ of the hyperfine effect in $B_d$. At the same time the
correction to the width ratio stemming from the kinetic operator must be
rather insignificant. The main uncertainty in the ratio 
$\Gamma_{\rm sl}(\Omega_b)/\Gamma_{\rm sl}(B_d)$ is associated with the Darwin
expectation value which has not yet been analyzed in detail for the state.
Assuming, in the spirit of the most naive nonrelativistic quark models, that
the underlying difference with $\lab$ is mainly related to the larger $s$-diquark
mass while its spin playing a secondary role, 
one would expect $\rho_D^3$ in $\Omega_b$ to exceed that in $\Lambda_b$. In
this case $\Omega_b$ may turn out to have the semileptonic width below that of
$B_d$ and, potentially, the smallest among all $b$ particles.

The prospects for studying the lifetime of $\Omega_b$ at the LHC are more
optimistic since the machine should produce a sufficiently large data sets
for it. The theoretical predictions for the corrections to the nonleptonic
width are not as definite, however, being affected by the significant 
spectator-specific effects in the KM-allowed modes.

\section{Charmless semileptonic decays}

So far we have discussed the total widths in the CKM allowed channels $b\tto
c\,e\nu$ and $b\tto c\,\mu \nu$. The $b\tto c(u)\,\tau\nu_\tau$ channel and
the $b\tto u\,\ell\nu$ decays are usually considered independently.  
The effect of the $\tau$ mass in $b\tto c\,\tau\nu_\tau$ does not qualitatively
change the predictions, therefore adding it would affect little the overall 
ratios of the semileptonic widths.

The $b\tto u\,\ell\nu$ channel is generally suppressed by a factor
$2|V_{ub}/V_{cb}|^2\!\sim \! 0.02$; including it in $\Gamma_{\rm sl}$ would modify
the width ratios at the per mil level. Consequently it is more relevant to
discuss the ratios of the $b\tto u\,\ell\nu$ widths directly, for $B_d$, $B_s$
and $\lab$.

The effect of the kinetic operator is fully universal and does not depend on
the channel at all. The chromomagnetic coefficient depends, strictly speaking,
on the quark mass in the final state, but this dependence is not
spectacular: the charm mass enhances the chromomagnetic effect compared to the
$b\tto u$ case by a factor about $1.35$. Therefore, one obtains here 
the same small effect as for the $b\tto c \,\ell\nu$ widths. 

The largest splitting comes from the Darwin operator which, 
for $b\tto u\,\ell\nu$ is 
entangled with the $SU(3)$-breaking between $B_d$ and $B_s$ 
in the non-valence WA. In $\lab$ the latter corresponds to what is
conventionally referred to as the PI mechanism. 
These actually comprise the
counterpart of the IC corrections discussed for $b\tto c \,\ell\nu$; in the
higher-order corrections only the `regular' terms suppressed by powers of
$1/m_b$ should be considered. 

A recent discussion of the valence and non-valence WA effects in $B$
mesons can be found in Ref.~\cite{icsieg}. Both were estimated as negative yet
quite small, around $-1.5\%$. Even though tentative, this suggests that the
actual effect in the $SU(3)$ breaking between $B_s$ and $B_d$ will be dominated
by the difference in the Darwin expectation value, 
\beq
\delta \, \frac{\Gamma_{\rm sl}(B_s\tto X_u\ell\nu)}
{\Gamma_{\rm sl}(B_d\tto X_u\ell\nu)} \approx -(1.5 \:\mbox{to}\:3)\% \,,
\label{112}
\eeq
with a totally negligible impact on the combined ($b\tto c\,\ell\nu$ and 
$b\tto u\,\ell\nu$) semileptonic width. 

Turning to $\Gamma_{\rm sl}(\lab \tto X_u\ell\nu)$ the previous reasoning is
amended by the potentially largest effect, the direct PI in $\lab$
\beq
\delta_{\rm PI}\, \frac{\Gamma_{\rm sl}(\lab\tto X_u\ell\nu)}
{\Gamma_{\rm sl}(B_d\tto X_u\ell\nu)} \simeq \frac{32\pi^2}{m_b^3}
  \,\frac{1}{2M_{\Lambda_b}} \matel{\lab}{\bar b \vec \gamma (1\!-\!\gamma_5)u\,
\bar u \vec \gamma (1\!-\!\gamma_5)b}{\lab}
\label{114}
\eeq
which -- in contrast to WA in $B$ -- is not prone to 
chirality suppression ab initio. 
Neglecting the non-valence $s$-quark effect (apparently suppressed)
Ref.~\cite{four} estimated the above expectation value to be about
$2\lambda'\!-\!\frac{1}{2}\lambda\!=\!0.025\GeV^3$. Adopting this we obtain
\beq
\delta_{\rm PI}\, \frac{\Gamma_{\rm sl}(\lab\tto X_u\ell\nu)}
{\Gamma_{\rm sl}(B_d\tto X_u\ell\nu)} \approx 8.5\%\, ,
\label{116}
\eeq
notably the dominant enhancement effect. The $b\tto u$ semileptonic width
difference with $B_d$ is then expected about $10\%$:
\beq
\frac{\Gamma_{\rm sl}(\lab\tto X_u\ell\nu)}
{\Gamma_{\rm sl}(B_d\tto X_u\ell\nu)} \simeq 1.1\,;
\label{118}
\eeq
this still may only slightly offset, by $0.2\%$ our estimate of the combined
semileptonic width of $\lab$ as compared to $B_d$.

\section{The semileptonic widths in the phase-space model}

The paper \cite{gr} let us look again into the numerics of the QCD-based HQE 
predictions for the splitting between the $b$-particle semileptonic decay widths.
The paper used a simple model for hadronization corrections derived solely 
from the phase-space effects in the decays into the lowest pseudoscalar and
vector final state mesons (or into the ground-state baryon, for 
the $\lab$ decays), which gave 
\beq
\frac{\Gamma_{\rm sl}(B_s)}{\Gamma_{\rm sl}(B_d)}=1.03, \qquad
\frac{\Gamma_{\rm sl}(\lab)}{\Gamma_{\rm sl}(B_d)}=1.13 \,.
\label{150}
\eeq
Not surprisingly, these differ from what the HQE predicts: it has been known
from the early 1990s that the phase-space--based calculations generally yield
power corrections scaling like $1/m_Q$ and these depend on $\La$, the energy
gap distinguishing the hadron mass from the heavy quark one. This parameter is
the principal nonperturbative quantity in differentiating the beauty hadrons
in question.  It is instructive to examine the model \cite{gr} from this
perspective; this allows to readily understand the numeric pattern in
Eq.~(\ref{150}).

The $b\tto c\,\ell\nu$-mediated decays is the most transparent case where 
the expansion in both $1/m_b$ and in $1/m_c$ applies. One only needs to employ
the hadron mass expansion
\bea
\nonumber
M_B \msp{-4}& = & \msp{-4} m_b+\La +\frac{\mu_\pi^2\!-\!\mu_G^2}{2m_b} + 
{\cal O}\!\left(\!\frac{1}{m_b^2}\!\right) +...\,, \qquad
M_{B^*}= m_b+\La +\frac{\mu_\pi^2\!+\!\frac{1}{3}\mu_G^2}{2m_b} + 
{\cal O}\!\left(\!\frac{1}{m_b^2}\!\right) +...\,,\\
M_{\lab} \msp{-4}& = & \msp{-4} m_b+\La_\Lambda +\frac{\mu_\pi^2(\lab)}{2m_b} + 
{\cal O}\!\left(\!\frac{1}{m_b^2}\!\right) +...
\label{152}
\eea
and likewise for charm. The corrections to the decay width assuming the 
absence of the 
formfactors and the $3\!:\!1$ ratio of the $D^*$ to $D$ yields (the adopted
assumptions in Ref.~\cite{gr}) then take the form
\beq
\Gamma_{\rm sl}(H_b)\simeq 
\Gamma_{\rm sl}(b)\left[1+C_{\bar \Lambda}\,\frac{\La}{m_b} + 
C_\pi \, \frac{\mu_\pi^2}{2m_b^2}+  C_G \, \frac{\mu_G^2}{2m_b^2}+...
\right],
\label{154}
\eeq
a series extended to any desired level. In particular, there is a strong
dependence on $\La$, $C_{\bar \Lambda}\! \simeq\!1.77$ that would dominate the width
differences. The total width in such a model as a function of $\La$ is well 
approximated by this linear dependence. 

Using $\La_\Lambda\!-\!\La\!\simeq M_{\lab}\!-\!\bar M_B\!\simeq\!300\MeV$
Eq.~(\ref{154}) would yield
\beq
\delta_{\La}\,  \frac{\Gamma_{\rm sl}(\lab)}{\Gamma_{\rm sl}(B_d)}
\simeq 11.5\% \, ;
\label{156}
\eeq
i.e., it reproduces the bulk of the Ref.~\cite{gr} prediction of $13\%$. 

At first glance, there is something strange with the smaller
difference for $B_s$ where scaling the above estimate by the ratio of 
$\delta \La$ one would estimate 
\beq
\delta\, \frac{\Gamma_{\rm sl}(B_s)}{\Gamma_{\rm sl}(B_d)} \approx 
\frac{M_{B_s}\!-\!M_B}{M_{\lab}\!-\!M_B}\cdot 
\delta \,\frac{\Gamma_{\rm sl}(\lab)}{\Gamma_{\rm sl}(B_d)}\approx 
\frac{87\MeV}{339\MeV}\cdot 0.13 \approx 3\%
\label{158}
\eeq
vs.\ $1.2\%$ obtained in Ref.~\cite{gr}. A closer look reveals the origin of
the reduction from $3\%$ to $1.2\%$: it is an effect of the kinetic term in
Eq.~(\ref{154}). In the phase-space approximations  the
coefficient $C_\pi$ typically is very large. 
Here at $m_c\!=\!1.25\GeV$ and  $m_b\!=\!4.6\GeV$ it comes out  
\beq
C_\pi \approx -15 \; ; 
\label{160}
\eeq
on the contrary, the OPE ensures the universal value of $c_\pi\!=\!-1$
regardless of the underlying dynamics. Using the value in Eq.~(\ref{54}) 
for the $\mu_\pi^2$
splitting -- that would stem from the literal comparison of the meson masses --
the inflated $C_\pi$ value in Eq.~(\ref{160}) yields a $-1.5\%$ correction to
be added to the $\La$ effect in  Eq.~(\ref{158}). (We can note
that for some partially accidental reason the effect of the chromomagnetic
term $C_G$ comes out approximately correct for the semileptonic width in the case of
$V\!-\!A$ weak interaction \cite{versus,bs} provided the $D^*$ to $D$ yield is
taken $3\!:\!1$ canceling the phase space effect in the final state.)

Thus we conclude that the moderate size of the correction in
$\Gamma_{\rm sl}(B_s)$ in the phase-space model comes as a result of a
cancellation between the $\La/m_b$ term and a $15$-fold inflated effect of the
kinetic operator. Neither are in the widths in reality. 

The above heavy mass expansion has certain peculiarities in the case of $b\tto
u\,\ell\nu$ widths.  Here Eq.~(\ref{154}) takes the form
\beq
\Gamma_{\rm sl}(H_b\tto X_u \ell\nu)\simeq 
\Gamma_{\rm sl}(b\tto u \,\ell\nu)\cdot
\left[1+5\,\frac{\La}{m_b} -8\frac{\sum w_i M_i^2}{m_b^2}+ 
\left(\!10\frac{\La^2}{m_b^2} \!+\!\frac{5\mu_\pi^2}{2m_b^2}\!\right)+...
\right]
\label{154u}
\eeq
where $M_i$ are the meson masses in the final state and $w_i$ the
corresponding branching fractions. 
In QCD the equivalent of the sum $\sum w_i M_i^2$ equals to $\frac{5}{8}\La
m_b$; the exact factor actually depends on the kinematics precisely in such a
way that the second and the third term cancel at arbitrary $q^2$ of the lepton
pair. However, the phase-space model of Ref.~\cite{gr} considers only the
lowest pseudoscalar and vector mesons with light quarks; then the third term
inevitably scales like $1/m_b^2$. As a result, the $5\La/m_b$ correction
remains unabated, at least for sufficiently heavy decaying hadron.

It is nevertheless interesting to note that the model intrinsically includes
an effect of WA in mesons estimated to be about $-2\%$ for the $b\tto
u\,\ell\nu$ rates.  No mechanism for the annihilation proper was actually
introduced, and the whole effect belongs to the realm of the `generalized WA'
phenomenon put forward in Ref.~\cite{WADs}, related, in particular, to the
annihilation-driven upward shift in the pseudoscalar masses. This readily
explains the negative sign of the `annihilation' correction to the width. The
thus generated WA is rooted in the QCD's $U(1)$-problem and is expressed
through the anomalous mass square of the $\eta'$ meson.

At the same time, the WA in the model of Ref.~\cite{gr} effectively scales
like $1/m_b^2$ whereas in reality it must be $1/m_b^3$. The reason is that all
the $b\tto u$ width in the model is attributed to the lowest $\pi$ and $\rho$
states, together with their unitary siblings. On the contrary, in QCD these
individual states are responsible for only a $1/m_b^3$ fraction of the decay
probability. Another distinct kinematic feature is that the related
flavor-dependent corrections are present for all kinematics, including fast
$\pi_0, \eta, \eta'$ with $|\vec p|\!\approx\!\frac{m_b}{2}$.  The leading
contribution to WA in QCD originates from the domain of low-momentum hadronic
final states.

Finally, let us note that another WA-like mechanism -- the PI in the $\lab
\tto X_u\ell\nu$ decays -- appears to be the largest preasymptotic correction
in beauty, unrelated to any mass shift. No room for such effect is seen in the
phase-space models.

\subsection{On the $\La$ effects}

The main fact of the HQE in QCD is the absence of the
$\La/m_b$ corrections in the inclusive widths, which underlies the principal
numerical difference with the phenomenological models that emphasize the
phase-space effects. In particular, the latter are enhanced by the
large power of the heavy quark mass commanding the partonic width, that
naively does not show up in the transition amplitudes. This aspect of the
QCD result was elucidated in Ref.~\cite{five}. 

Comparing the actual QCD with the phase-space model, notably for the
$b\tto c\,\ell\nu$ transitions where both $b$ and $c$ quarks are treated 
as heavy, one
observes that the hadronic transition amplitudes into $D$ and $D^*$ are
additionally suppressed nonperturbatively, at nonvanishing recoil, by the
corresponding formfactors. These effects at a given velocity do not fade away
even for infinitely heavy quarks. This would have led to the mismatch between
the partonic and hadronic calculations already at a ${\cal O}(1)$ level. As
noted in the mid 1980s by Voloshin and Shifman \cite{svwidth}, this is eliminated
by the production of the truly excited charmed states. In the SV regime these
are the various $P$-wave states, and the cancellations of the $1/m_Q^0$
corrections in the SV limit is expressed by the Bjorken sum rule
\cite{bjorken}
\beq
\varrho^2\!- \mbox{$\frac{1}{4}$}= 
\sum_n 2|\tau^{(n)}_{3/2}|^2 +\sum_n  |\tau^{(n)}_{1/2}|^2.
\label{172}
\eeq

$P$-waves are not included in the phase-space model of  Ref.~\cite{gr}, yet
neither are the formfactor effects which should amount to setting the IW
slope $\varrho^2\!=\!\frac{1}{4}$ in $B$ (or to zero in $\lab$). The
precise shapes assumed for the formfactors are actually unimportant in the
calculations of the widths ratios at this leading level, only that 
they are the same. 

That becomes less obvious for the terms $\La/m_b$. It was detailed
in Ref.~\cite{optical} that the absence of such corrections in the widths
comes from the sum rule for the first moment of the structure functions; in
the SV regime this becomes the `optical' sum rule from Voloshin \cite{volopt}:
\beq
\frac{\La}{2}=\sum_n  2\varepsilon_n |\tau^{(n)}_{3/2}|^2 + 
\sum_n \varepsilon_n |\tau^{(n)}_{1/2}|^2
\label{174}
\eeq
including, again, only the final states with the $P$-wave quantum numbers. The
relation, however is more general and ensures vanishing of the $\La/m_b$
effects for the arbitrary final state quark mass. 

There have been later analyses focusing on the way the cancellation occurs
upon combining the ground-state and the $P$-wave state yields in $B$-decays in
the SV regime. In particular, Ref.~\cite{intsumrul} scrutinized a more
complicated case of the axial-vector transitions.\footnote{A.\,Vainshtein had 
presented a similar sum rule, but never published it.} A dedicated 
review presentation can be found in Ref.~\cite{bigmanyellow}.

Now the inherent problem of the phase-space models should become evident:
lacking a dynamic description of the transition formfactors and of the $P$-wave
amplitudes they have to assume the structureless ground-state formfactors and 
absent $P$-wave transitions, to save duality to the leading order in
$1/m_Q$. However, this only pushes the problem to the next order
$1/m_Q$. Once the $P$-wave amplitudes are absent, there may be no difference
between the quark and hadron kinematics: $\La$ has to vanish. Adopting the
absence of the $P$-wave transitions and non-zero $\La$ simultaneously can not
be reconciled in any quantum mechanical system.

\section{Conclusions}

We have presented our expectations for the ratios of the semileptonic decay
rates of $B_d$, $B_s$ and $\Lambda_b$ hadrons applying the OPE-based $1/m_Q$
expansion. In $\Gamma_{\rm sl}(B_s)/\Gamma_{\rm sl}(B_d)$ the effect
originates from the $SU(3)$ breaking in the heavy quark expectation
values. While generally the $SU(3)$ splittings may be expected to constitute
$30$ to $50\%$, we found a peculiar pattern: the leading terms in the
expansion appear to be far more $SU(3)$-robust, especially those describing
the spin effects. The full-strength $SU(3)$ breaking apparently is delayed to
higher orders in $1/m_Q$, thereby suppressing the overall $SU(3)$ violating
effects in the beauty mesons. 

We expect a slower semileptonic decay of $B_s$ mesons compared to
$B_d$, by about $1\%$, mainly due to a larger Darwin expectation value
$\rho_D^3(B_s)$. Other effects derived from the observed heavy meson masses
appear to be notably smaller. 

The prediction  $\delta \Gamma_{\rm sl}(B_s)/\Gamma_{\rm sl}(B_d)\approx -1\%$
is tied to the expected larger value for the Darwin operator in
$B_s$. This leads to other implications, for instance a larger slope of the IW
formfactor for the strange meson, and calls for the independent verifications.  

This finding is peculiar from a certain perspective: the natural scale for the
$SU(3)$ breaking in the individual exclusive channels is generally described
by a parameter like 
$f_K^2 / f_\pi^2 \!\approx\! 1.4$; this is much larger than what is seen
in the inclusive rates. Therefore, the duality with the OPE description
assumes nontrivial cancellations between the various exclusive channels to
ensure only a small $SU(3)$ breaking in their sum.

No symmetry promotes equality of the decay rates in $\lab$ and $B$.
Nevertheless the OPE predicts only a small difference. The largest effect,
about $3\%$ is the absence in $\lab$ of the suppression from the
chromomagnetic interaction. A potentially significant contribution to the
width difference from the Darwin operator appears suppressed according to the
dynamics based estimated.

In $b\tto u\,\ell\nu$ semileptonic widths the deviation from unity of 
the $B_s$ to $B_d$ ratio 
is expected to be small, $\delta \Gamma(B_s\tto
X_u\ell\nu)/\Gamma(B_d \tto X_u\ell\nu)\!  \approx \!-2.5\%$, although this
number reflects somewhat model-dependent estimates \cite{icsieg} of the
non-valence WA. At the same time a similar effect -- the PI in $\lab$ in the
$b\tto u\,\ell\nu$ channel -- emerges as the largest effect driving up the
ratio to
$$
\frac{\Gamma(\lab\tto X_u\ell\nu)}{\Gamma(B_d \tto X_u\ell\nu)} \approx 1.1\,.
$$

In an even more remote perspective, we have found semileptonic 
$\Omega_b$ decays to be of a particular interest. Its semileptonic 
width is expected to be slightly
suppressed and close to that of $B_d$ meson; it may even turn out the smallest
among all $b$ hadrons -- in spite of the remarkably large mass of $\Omega_b$. The
uncertainty in the OPE prediction for 
$\Gamma_{\rm sl}(\Omega_b)/\Gamma_{\rm sl}(B_d)$ may be 
reduced once their masses along
with the hyperfine partner about $23\MeV$ higher are accurately measured,
and if the information on their $P$-wave states in the beauty or charm 
sectors becomes available.

The OPE-based predictions appear quite different from the phase-space
models. For $B_s$ we expect a smaller effect of the opposite overall sign; in
$\Gamma_{\rm sl}(\lab)$ the phase-space model cannot avoid a much larger effect
of about $13\%$ due to the term scaling like $\La/m_b$. In the $b\tto u\,\ell\nu$
channel the significant effect, up to $10\%$ comes in the OPE from the
flavor-dependent processes involving the interference with the 
spectator quark, a mechanism having no counterpart in the phase-space picture. 

The ideas to reduce the bulk of the hadronization effects in the inclusive
heavy flavor decays to the phase-space corrections alone have a long history.
Perhaps the best known was paper \cite{altmart} which sought to explain the
existed significant experimental difference in the $\lab$ vs.\ $B_d$ lifetimes
by overriding the OPE results with the assumed $M_{H_Q}^5$ scaling of the
widths. The evident fact that baryons in the decay products of $\lab$ have
likewise larger mass than their meson counterparts, was discarded. This
particular drawback has been eliminated in Ref.~\cite{gr}.  Accounting for the
increase in the final-state mass reduced, as expected, the shift in the resulting
(semileptonic) decay rate ratio 
$\Gamma_{\rm sl}(\Lambda_b)/\Gamma_{\rm sl}(B_d)$ from $1.37$ to $1.13$, 
a value much closer to the OPE expectation
-- yet still markedly above it.

\subsection*{Acknowledgments}

We thank A.\,Khodjamirian for useful discussions and S. Turczyk for help with
evaluating the estimated higher-order power effects. The communication from
M.\,Artuso raising the problem in question is acknowledged.  The study enjoyed
a partial support from the NSF grant PHY-0807959 and from the
RSGSS-65751.2010.2 grant. TM is partially supported by the German research
foundation DFG under contract MA1187/10-1 and by the German Ministry of
Research (BMBF), contracts 05H09PSF.

\end{document}